\documentclass[rsi,
 amsmath,amssymb,
 reprint,%
]{revtex4-1}

\usepackage{graphicx,dcolumn,bm,mathptmx,etoolbox,bbold}
\usepackage[hidelinks]{hyperref}
\usepackage[utf8]{inputenc}
\usepackage[T1]{fontenc}
\usepackage{placeins}
\makeatletter
\def\@email#1#2{%
 \endgroup
 \patchcmd{\titleblock@produce}
  {\frontmatter@RRAPformat}
  {\frontmatter@RRAPformat{\produce@RRAP{*#1\href{mailto:#2}{#2}}}\frontmatter@RRAPformat}
  {}{}
}
\makeatother
\begin{document}

\preprint{AIP/123-QED}

\title[Sample title]{Fast and Automated Optical Polarization Compensation of Fiber Unitaries}
\author{Niklas Braband}
\affiliation{Department of Physics, University of Erlangen-Nuremberg, Staudtstraße 7, 91058 Erlangen, Germany}
\author{Arman Mansouri, Riza Fazili, Stefanie Czischek, and Jeff S. Lundeen*}
\affiliation{Department of Physics and Nexus for Quantum Technologies, University of Ottawa, 25 Templeton Street, Ottawa, Ontario, Canada K1N 6N5}
\email{jlundeen@uottawa.ca}

\begin{abstract}
The polarization of light is critical in various applications, including quantum communication, where the photon polarization encoding a qubit can undergo uncontrolled changes when transmitted through optical fibers. Bends in the fiber, internal and external stresses, and environmental factors cause these polarization changes, which lead to errors and therein limit the range of quantum communication. To prevent this, we present a fast and automated method for polarization compensation using liquid crystals. This approach combines polarimetry based on a rotating quarter-waveplate with high-speed control of the liquid-crystal cell, offering high-fidelity compensation suitable for diverse applications. Our method directly solves for compensation parameters, avoiding reliance on stochastic approaches or cryptographic metrics. Experimental results demonstrate that our method achieves over 99\,\% fidelity within an average of fewer than six iterations, with further fine-tuning to reach above 99.5\,\% fidelity, providing a robust solution for maintaining precise polarization states in optical systems.
\end{abstract}

\maketitle

\section{\label{sec:I}Introduction}

The polarization of light plays a crucial role in a diverse range of applications, such as quantum cryptography, medical and biological imaging \cite{Liao2017, He2021, Intaravanne2022}. However, the polarization state can significantly alter as the light passes through various media, particularly in optical telecommunication fibers \cite{czegledi_2016}. In these fibers, the routing geometry (e.g., bending or twisting) and manufacturing imperfections lead to birefringence, wherein the orientation of the slow and fast axes and the degree of birefringence vary along the fiber. In turn, this birefringence causes an unpredictable change in the input state of polarization (SOP) and introduces errors in the information encoded in polarization.

For example, an input bit might be encoded as horizontal (0 bit value) or vertical polarization (1 bit value) but the output might be right- or left-handed circularly polarized. Since the latter is an equal combination of horizontal and vertical polarization, the information is effectively completely randomized. This deleterious effect is compounded by its variability over time, influenced by movement and environmental changes such as temperature and pressure. The resulting error rate is particularly a problem for quantum key distribution (QKD) \cite{Neumann2022,Tan2024}, since when combined with photon loss and detector dark counts, it limits the range over which the cryptographic key can be transmitted. Polarization manipulation is also essential in various areas, with applications ranging from biomedical imaging to astronomical observation \cite{li_high-sensitivity_2024,he_vectorial_2023}. Therefore, continuous polarization monitoring and rapid polarization manipulation would help maintain consistent performance in technologies reliant on accurate polarization states.

A straightforward method for characterizing an unknown polarization state is the rotating quarter-waveplate (QWP) technique \cite{Goldstein_polarization}. This approach uses a quarter-wave plate and a polarizer to measure the four-element Stokes vector. By rotating the waveplate and analyzing the resultant light intensity changes, all components of the Stokes vector are determined. The speed of this method can vary depending on the rotation frequency of the waveplate, with faster rotation increasing the characterization speed. Once the SOP is known, a rapid and arbitrary polarization transformation can be achieved using three liquid crystal cells. This allows for the transformation of the SOP to a desired state, controlled solely by adjusting the operating voltages of the liquid crystals \cite{Zhuang1999}. 

Existing polarization compensation methods often rely on stochastic techniques, interferometry, or quantum-bit-error-rate (QBER) minimization, which can be inefficient, particularly in long-range QKD systems where they result in low key rates \cite{shi_fibre_2021,Stromberg:24,ramos_full_2022,Tan2024, Peranic2023, Bedroya2024}. In Ref. \cite{Dai:19}, active compensation of polarization errors is achieved with adaptive optics via feedback to the polarization state generator, which is impractical in applications with limited control of the initial polarization. The work presented in \cite{GRAMATIKOV2020163474} employs a polarimetry technique using a single liquid crystal cell, where both the voltage and angle are varied. However, the intermittent rotation of the liquid crystal, with frequent stops, poses a challenge to achieving fast and reliable compensation. Additionally, the work lacks a rigorous proof of the ability to change an arbitrary polarization to another, as demonstrated in \cite{Zhuang1999}.

In this manuscript we introduce a novel approach that relies on polarization tomography and directly solving for the compensation parameters, enabling fast polarization compensation with high fidelity. Specifically, we integrate the rotating quarter-waveplate polarimetry technique with polarization transformation by adjusting three liquid crystal cell voltages. Since our method does not rely on cryptographic figures of merit, it can be implemented on polarization-based QKD systems regardless of the range and key rates. Contrary to \cite{GRAMATIKOV2020163474}, we only vary the voltage of three liquid crystals set at fixed angles, quickly compensating for the effect of any arbitrary unitary transformation. 

Moreover, there is a lack of comprehensive experimental compensation procedures, including required codes from detection to correction of polarization, which has hindered their broader adaptation. To address this, our work includes detailed technical information and essential codes, offering a helpful resource for those interested in implementing and exploring these methods. We believe this work will be beneficial in many systems requiring precise and continuous polarization control.

\begin{figure}[!ht!]
    \centering
    \includegraphics{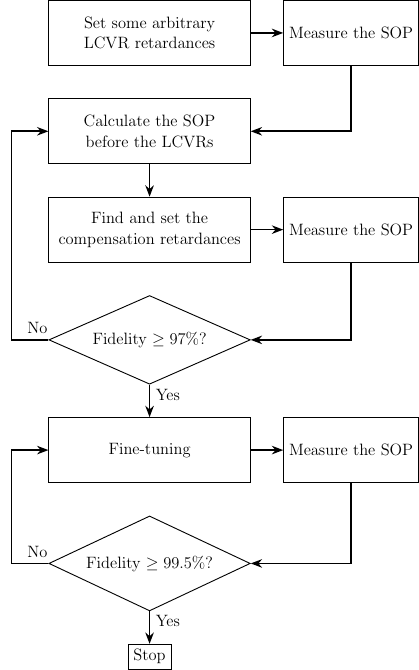}
    \caption{Algorithm for compensation of fiber unitaries. Here, SOP and LCVR stand for the state of polarization and liquid crystal variable retarders, respectively. We propose and implement this algorithm, which relies on polarization tomography and directly solving for the liquid crystal compensation retardances. Once a fidelity threshold is crossed, small variations of the retardances are used to reach the desired fidelity (here was 99.5\,\% fidelity).}
    \label{fig:flowchart}
\end{figure}

A flowchart of our compensation scheme is presented in Fig.~\ref{fig:flowchart}. In Section II, we provide a brief overview of the theoretical framework for polarization and its transformation, focusing on Stokes vectors and Mueller matrices. Section III presents the rotating QWP tomography method. In Section IV, we discuss the polarization transformation and characterization of liquid crystals, along with the compensation algorithm and results obtained using a continuous-wave (CW) 808\,nm laser (bandwidth of $\sim$ 1\,nm). Our automated, direct compensation algorithm reaches the desired fidelity thresholds within a few iterations, with speed primarily governed by the QWP rotation rate.

\section{\label{sec:II}Stokes parameters and Mueller matrices formalism}

Light, being a transverse electromagnetic wave, exhibits polarization — a fundamental characteristic that plays a crucial role in its interaction with materials and devices. The formalism necessary to represent the polarization of light and its manipulation is presented in this section, with comprehensive treatments being available in \cite{Gil2007, Perez2022}. A wave propagating in the $z$-direction can be described with two orthogonal electric field components having amplitudes $E_{x,0}$ and $E_{y,0}$, and phases $\delta_x$ and $\delta_y$ \cite{hechtoptics,Goldstein_polarization}:

\begin{align}
    E_x(z,t) &= E_{x,0} \cos(\omega t - k z + \delta_x), \\
    E_y(z,t) &= E_{y,0} \cos(\omega t - k z + \delta_y),
\end{align}

\noindent where $\omega$ is the angular frequency and $k$ is the angular wave number. The Stokes parameters are defined as follows:

\begin{align}
S_0 &= E_{x,0}^2 + E_{y,0}^2, \\
S_1 &= E_{x,0}^2 - E_{y,0}^2, \\
S_2 &= 2E_{x,0} E_{y,0} \cos(\delta), \\
S_3 &= 2E_{x,0} E_{y,0} \sin(\delta),
\end{align}
where $\delta = \delta_y - \delta_x$. They are usually written as a vector, called the Stokes vector $\vec{S}$:
\begin{equation}
\vec{S}=\begin{bmatrix}
S_0 \\
S_1 \\
S_2 \\
S_3 \\
\end{bmatrix} = S_0 \begin{bmatrix}
    1 \\ \mathsf{S_1} \\ \mathsf{S_2} \\ \mathsf{S_3}
\end{bmatrix},  \label{eq:Stokes-vector}
\end{equation}

\noindent where $S_0$ describes the intensity of the wave, $S_1$ the intensity difference between horizontally (H) polarized light and vertically (V) polarized light, $S_2$ the intensity difference between diagonally (D) polarized light and anti-diagonally (A) polarized light, and $S_3$ the intensity difference between right-hand (R) circularly polarized light and left-hand (L) circularly polarized light. In Eq.~\eqref{eq:Stokes-vector}, $\mathsf{S}_i = S_i/S_0$ represent the normalized Stokes parameters. The three-component vector $\vec{\mathsf{S}}=(\mathsf{S}_1, \mathsf{S}_2, \mathsf{S}_3)$ thus describes the Cartesian coordinates of the polarization state in the Poincare sphere.

In light sources where the polarization changes very rapidly and irregularly, the observed polarization state is partially polarized or unpolarized (e.g., for randomly oriented atomic emitters). A Stokes vector can describe light with any degree of polarization, with $S_0$ being related to the other parameters via the relation:
\begin{equation}
    S_0^2 \geq S_1^2 + S_2^2 + S_3^2.
    \label{eq:dop}
\end{equation}
For completely polarized light, Eq.~\eqref{eq:dop} reduces to an equality. 
\newpage 
The polarization of light can be changed using optical elements such as retarders, polarizers, or depolarizers. These changes to the Stokes parameters can be described by a Mueller matrix $\mathbf{M}$ with elements $m_{ij}$ as:
\begin{align}
\vec{S}'=\begin{bmatrix}
S_{0}^{'}\\
S_{1}^{'} \\
S_{2}^{'}\\
S_{3}^{'}
\end{bmatrix} = 
\begin{bmatrix}
m_{00} & m_{01} & m_{02} & m_{03}\\
m_{10} & m_{11} & m_{12} & m_{13}\\
m_{20} & m_{21} & m_{22} & m_{23}\\
m_{30} & m_{31} & m_{32} & m_{33}\\
\end{bmatrix}\cdot  
\begin{bmatrix}
S_{0}\\
S_{1} \\
S_{2}\\
S_{3}
\end{bmatrix}=\mathbf{M}\cdot \vec{S}.
\end{align}

For a system of $n$ optical elements the Mueller matrices can be multiplied to get the Mueller matrix describing the entire system:
\begin{align}
    \vec{S}'=\mathbf{M}_{n}\cdot ...  \cdot \mathbf{M}_{2}\cdot \mathbf{M}_{1}\cdot \vec{S}= \mathbf{M}_{\text{sys}}\cdot \vec{S}.
\end{align}

The fidelity $f(\rho, \sigma)$ is a measure of the closeness between two states $\rho$ and $\sigma$ and can be defined as \cite{Jozsa1994, Walls2008}:
\begin{equation}
    f(\rho,\sigma) = \left[\text{Tr}\sqrt{\sqrt{\rho}\sigma \sqrt{\rho}}\right]^2.
\end{equation}
In terms of Stokes vectors of SOPs $\vec{S}_a$ and $\vec{S}_b$, this can be expressed as \cite{Jozsa1994}:
\begin{align}
    &f(\vec{S}_a, \vec{S}_b) \nonumber \\
    &= \frac{1}{2}\left[1 + \sum^3_{i=1} \mathsf{S}_{i, a}\mathsf{S}_{i, b} + \sqrt{(1-|\vec{\mathsf{S}}_a|^2)(1-|\vec{\mathsf{S}}_b|^2)}\right]^2.
\end{align}
If one of the SOPs, for example $\vec{S}_b$ is fully polarized, $|\vec{\mathsf{S}}_b|^2=1$ and the fidelity reduces to:
\begin{equation}f(\vec{S}_{a},\vec{S}_{b})=\frac{1}{2}(1+\mathsf{S}_{1,a} \mathsf{S}_{1,b}+\mathsf{S}_{2,a} \mathsf{S}_{2,b}+\mathsf{S}_{3,a}  \mathsf{S}_{3,b}).
    \label{eq:fidelity}
\end{equation}
Given the target SOP that we try to match through polarization compensation is always fully polarized, we use Eq. \ref{eq:fidelity} to calculate the closeness between the measured and the target SOPs.
\section{\label{sec:III}Polarization tomography}

All Stokes parameters can be measured simultaneously using a rotating QWP, a polarizing beam splitter (PBS) and a photo detector (PD) or power meter \cite{Goldstein_polarization}. The incoming light first passes through the rotating QWP at an angle $\phi$ between the fast and the horizontal axes. The intensity $\tilde{S}_{0}$ transmitted by the PBS is then measured by the PD. The Stokes vector is thus transformed as follows:

\begin{widetext}
\begin{align}
&\vec{\tilde{S}}(\phi) =\mathbf{M}_{\text{PBS}}\cdot \mathbf{M}_{\text{QWP}}(\phi)\cdot\vec{S} \label{eq:rotQWP}=\frac{1}{2}\begin{bmatrix}
1 & 1 & 0 & 0 \\
1 & 1 & 0 & 0 \\
0 & 0 & 0 & 0 \\
0 & 0 & 0 & 0 
\end{bmatrix} \cdot \begin{bmatrix}
1 & 0 & 0 & 0 \\
0 & \cos^2(2\phi) & \sin(2\phi)\cos(2\phi) & -\sin(2\phi) \\
0 & \sin(2\phi)\cos(2\phi) & \sin^2(2\phi) & \cos(2\phi) \\
0 & \sin(2\phi) & -\cos(2\phi) & 0 
\end{bmatrix} \cdot
\begin{bmatrix}
S_{0}\\
S_{1} \\
S_{2}\\
S_{3}
\end{bmatrix}. 
\end{align}
\end{widetext}

\noindent From Eq.~\eqref{eq:rotQWP}, the transmitted intensity $\tilde{S}_{0}$ is given by:
\begin{align}
    I(\phi) &=\tilde{S}_{0}(\phi) \nonumber\\
&=\frac{1}{2} ( S_0+S_1 \cos^2 2\phi +S_2 \sin2\phi\cos2\phi-S_3 \sin2\phi)\nonumber\\
&=\frac{1}{2}(A_0+B_0\sin2\phi+C_0\cos4\phi+D_0\sin4\phi).
\end{align}

\noindent The intensity can be rewritten using 
\begin{align}
    A_0 &= S_0+\frac{S_1}{2},\ B_0 = -S_3,\ 
    C_0 = \frac{S_1}{2},\ D_0 = \frac{S_2}{2},
\end{align}
\noindent and can now be seen as a truncated Fourier series. As only a finite number of samples are measured for the intensity, each Fourier coefficient integral is turned into a sum with $\Delta\phi$ being the angle step size between two measurements and $N$ the number of samples, resulting in:

\begin{align}
A_0&=\frac{1}{\pi}\int_{0}^{2\pi} I(\phi) \,d\phi=\frac{2}{N}\sum_{n=1}^N I_n, \\
B_0&=\frac{2}{\pi}\int_{0}^{2\pi} I(\phi) \sin(2\phi)\,d\phi =\frac{4}{N}\sum_{n=1}^N I_n \sin(2n\Delta\phi),\\
C_0&=\frac{2}{\pi}\int_{0}^{2\pi} I(\phi) \cos(4\phi)\,d\phi=\frac{4}{N}\sum_{n=1}^N I_n\cos(4n\Delta\phi),\\
D_0&=\frac{2}{\pi}\int_{0}^{2\pi} I(\phi) \sin(4\phi)\,d\phi =\frac{4}{N}\sum_{n=1}^N I_n\sin(4n\Delta\phi).
\end{align}

\begin{figure}[h]
    \centering
    \includegraphics[height=2.5cm]{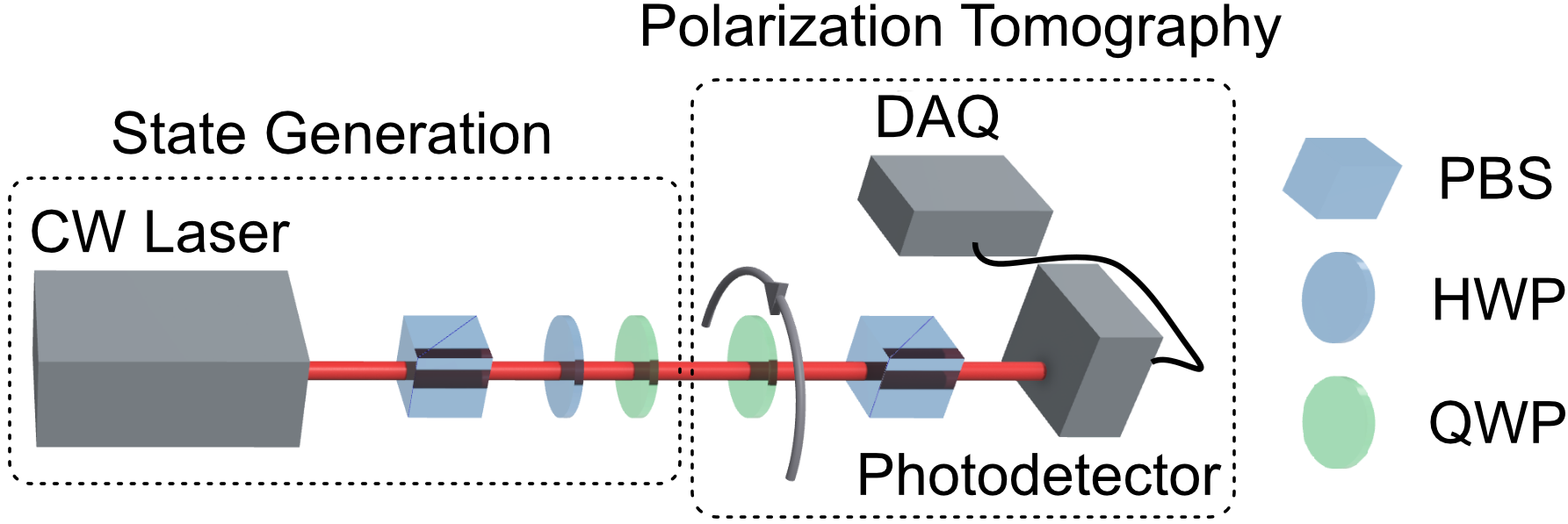}
    \caption{Polarization tomography setup consisting of a PBS to ensure an $\vec{S}_H$ state, a HWP and/or a QWP to prepare the different input states, and the SOP-measurement setup as described in Sec.~\ref{sec:III}. The photodetector voltages are recorded using a data acquisition device (DAQ).}
    \label{fig:tomography_setup}
\end{figure}

The setup in Fig.~\ref{fig:tomography_setup} is used to verify the polarimetric method. For the measurement, one full rotation of the QWP is used. The rotation mount used was rotating at its fastest possible rate, which is $30^\circ\,$s$^{-1}$. The position of the QWP was measured via the rotation mount with the highest possible sampling rate, resulting in an angle step size of approximately $1.16^\circ$. Accordingly, 310 data points were used for one polarization measurement. To prevent drift and ensure accurate angle measurements, the rotation mount was re-homed and calibrated after every fifth rotation.

\begin{figure}[h]
\centering
\includegraphics[width=0.47\textwidth]{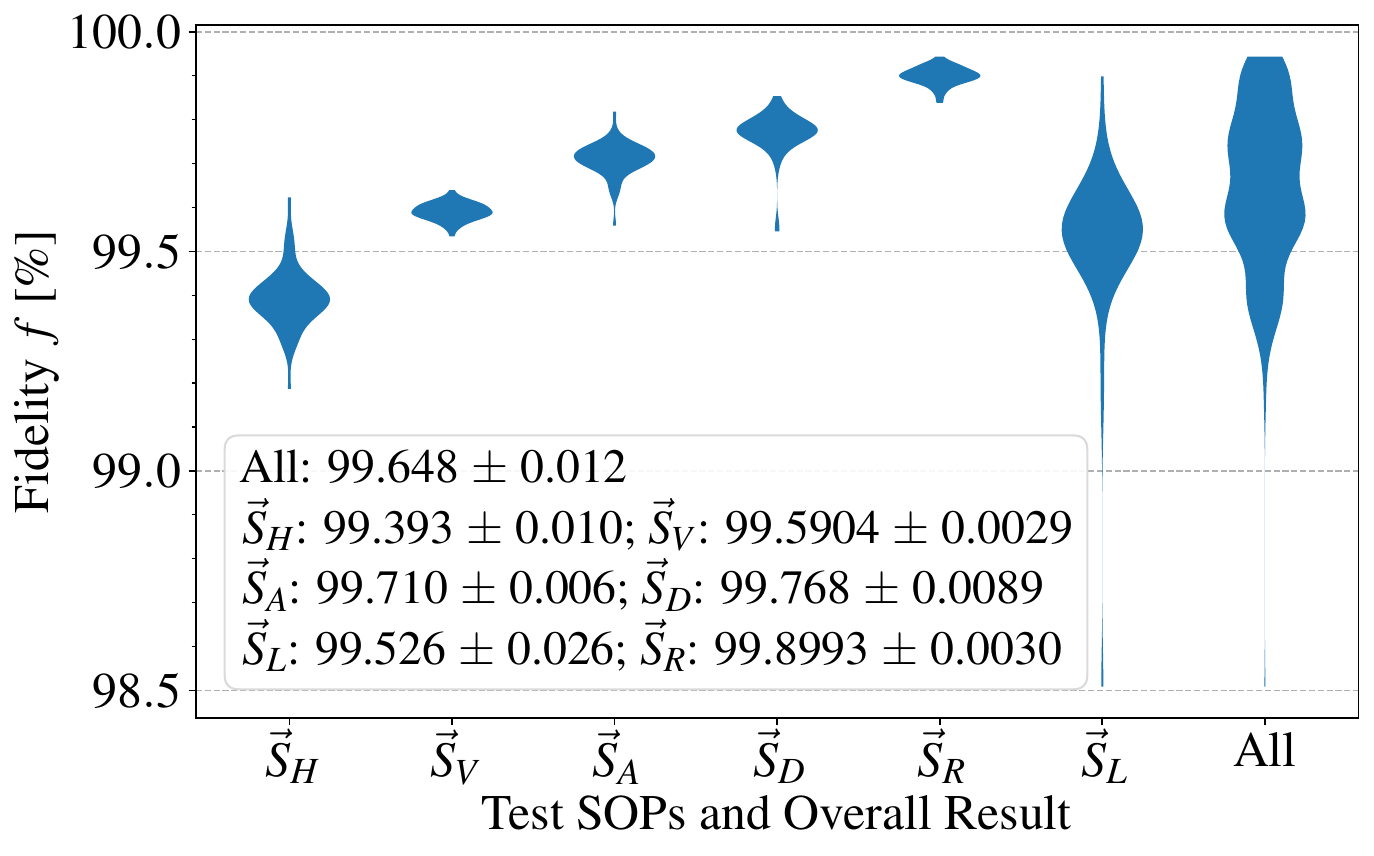}
\caption{Results of the polarization tomography for six different input states are shown as a violin plot. This plot shows the fidelity distribution of the 50 data points for each input state, along with the overall distribution, average fidelity, and standard error of the mean.}
\label{fig: Tomography_test}
\end{figure}

The polarization states $\vec{S}_{H}$, $\vec{S}_{V}$, $\vec{S}_{A}$, $\vec{S}_{D}$, $\vec{S}_{L}$, and $\vec{S}_{R}$ were prepared using a QWP and a half-wave plate (HWP) as test states for the tomography system. We performed multiple rounds of state tomography for these SOPs, and calculated their corresponding fidelities, as shown in Fig.~\ref{fig: Tomography_test}. The overall error for the polarimeter was estimated from the standard error in the mean of the measured fidelities of each SOP. Further details on the experimental setup can be found in the Appendix.

The polarization tomography achieves average fidelities of $99.65\,\%$ and relatively small statistical fluctuations of $0.012\,\%$ (Fig.~\ref{fig: Tomography_test}). The input laser power fluctuated by 0.1\% over a trial duration. The error in the fidelity due to these power fluctuations is estimated to be 0.004\% over 50 trials. The different mean fidelities for the input states are likely to be influenced by how well the specific input state was prepared. Therefore, the calculated standard error in the mean of the complete test measurement is most likely larger than the actual error of the polarimeter and thus a conservative value for the uncertainty.

\section{Polarization transformation}
\subsection{\label{sec:IV}Liquid crystal characterization}

Liquid crystal variable retarders (LCVRs) are molecules with directional properties that can be aligned to exhibit uniaxial birefringence \cite{Scott_LC}. By applying a square wave AC voltage, the retardance of the optical element changes with the root-mean square value of the applied voltage. The action of an LCVR can be described by the following Mueller matrix:
\begin{widetext}
\begin{align}
&\mathbf{M}_{\text{LCVR}}(\theta,\delta) =\begin{bmatrix}
1 & 0 & 0 & 0 \\
0 & \cos^2 2\theta+\sin^2 2\theta \cos \delta & \cos 2\theta \sin 2\theta (1-\cos\delta) & -\sin2\theta \sin\delta \\
0 & \cos2\theta \sin2\theta(1-\cos\delta) & \cos^2 2\theta \cos\delta+\sin^2 2\theta & \cos 2\theta \sin \delta \\
0 & \sin 2\theta \sin \delta & -\cos 2\theta \sin \delta & \cos \delta
\end{bmatrix}. 
\end{align}
\end{widetext}
Here, $\theta$ is the angle of the liquid crystal birefringent fast axis relative to the horizontal axis, and $\delta$ is the relative phase retardance between the horizontal and vertical polarizations (which varies with voltage).

A characterization of each LCVR is needed to obtain the relation between the applied voltage and the retardance at a given wavelength \cite{Schnoor:20}. This can be done using horizontally polarized input light $\vec{S}_{H}$. The light passes through a PBS, the LCVR at $45^\circ$, a HWP at $45^\circ$, and a second PBS. Both angles are measured between the fast axis of each component and the horizontal axis.

The Mueller matrix for a HWP with its fast axis forming an angle $\phi$ with the horizontal is given by:
\begin{widetext}
\begin{align}
\mathbf{M}_{\text{HWP}}(\phi)&=\begin{bmatrix}
1 & 0 & 0 & 0 \\
0 & \cos^2(2\phi)-\sin^2(2\phi) & 2\cos(2\phi)\sin(2\phi) & 0 \\
0 & 2\cos(2\phi)\sin(2\phi) & \sin^2(2\phi)-\cos^2(2\phi) & 0 \\
0 & 0 & 0 & -1
\end{bmatrix}.
\end{align}
\end{widetext}
The equations for $\vec{S}(\phi,\delta)$ and $I(\phi,\delta)$ are given by:
\begin{align}
\vec{S}(\phi,\delta)&= \mathbf{M}_{\text{PBS}} \cdot \mathbf{M}_{\text{HWP}}\left(\frac{\pi}{4}\right) \cdot \mathbf{M}_{\text{LCVR}}\left(\frac{\pi}{4},\delta\right) \nonumber \\
&\quad \cdot \mathbf{M}_{\text{PBS}} \cdot \mathbf{M}_{\mathrm{HWP}}(\phi)\cdot \vec{S}_{\text{H}},\\
I(\phi,\delta)&=S_{0}(\phi,\delta)=\frac{1}{2} \cos^2(2\phi)\left (1-\cos(\delta)\right).
\end{align}

The maximum intensity is reached when $\cos(\delta)=-1$:
\begin{align}
I_{\text{max}}&=\cos^2(2\phi), \\
I(\delta)=\frac{I_{\text{max}}}{2}(1-\cos(\delta)) &\Leftrightarrow \delta=\arccos\left(1-\frac{2I}{I_\text{max}}\right).
\label{eq: retardance}
\end{align}
This establishes the retardance-intensity relationship for a given LCVR voltage. By sweeping over the full voltage range, the complete retardance characterization curve as a function of voltage can be obtained.

Because of the arccosine relation between the intensity and the retardance, the result needs to be unwrapped. The final result should be a smooth curve with no jumps or sudden changes in the derivative. A detailed description of the phase unwrapping can be found in \cite{Lopez-Tellez}.

\begin{figure}[h!]
    \centering
    \includegraphics[scale=0.42]{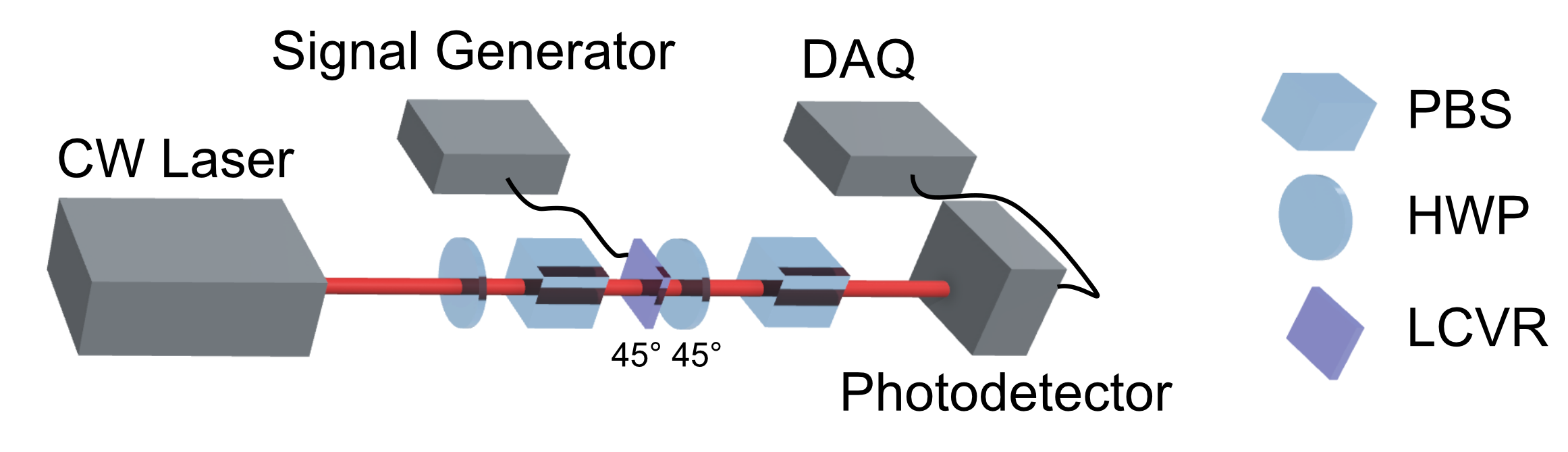}
    \caption{Setup to characterize liquid crystal variable retarders.}
    \label{fig:Char_setup}
\end{figure}

\begin{figure}[h!]
    \centering
    \includegraphics[width=8cm]{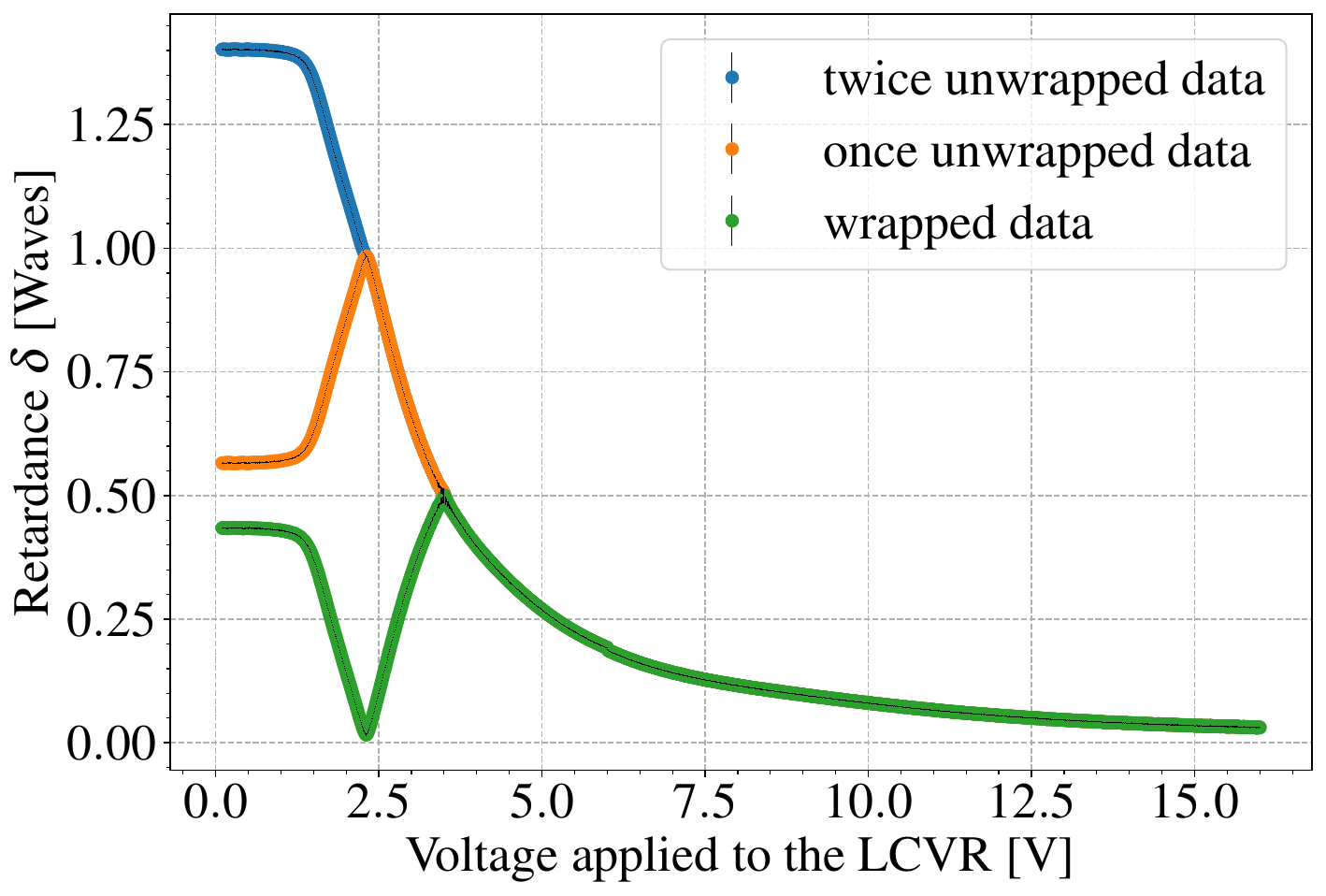}
    \caption{Example of the retardance-voltage relation of a LCVR at $\lambda$=810\,nm; raw data (green) was unwrapped twice; errors calculated via the Gaussian error propagation (see appendix for details).}
    \label{Characterization}
\end{figure}

The setup for the LCVR characterization is shown in Fig.~\ref{fig:Char_setup}. One would need to characterize each LCVR for the wavelengths used. We used compensated full-wave compensated LCVRs to access the full range of required retardances. During the characterization, a square wave voltage with a frequency of 2\,kHz was applied to the LCVR. The RMS voltage of the square-wave was sweeped from 0.1\,V to 16\,V with a step size of 0.01\,V. Further details on the experimental setup and the error calculation can be found in the appendix.

As shown in Fig.~\ref{Characterization}, unwrapping the raw data leads to the expected curve for the nonlinear retardance-voltage relation of an LCVR \cite{Scott_LC,Schnoor:20,Lopez-Tellez,Gladish_2010}. Because the final curve exceeds a full-wave of retardance ($2\pi$), two unwrapping steps are needed. As the retardance range of the LCVR is at least a full-wave, any possible retardance can be accomplished.

\subsection{\label{sec:level2}Polarization compensation}

\begin{figure}[h!]
    \includegraphics[scale=0.99]{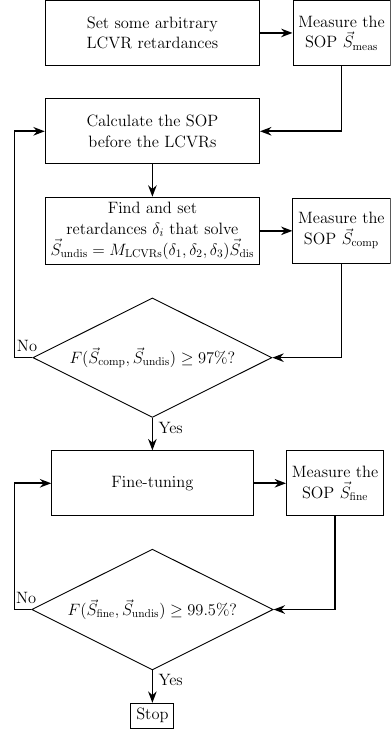}
    \caption{Detailed algorithm for the polarization compensation of fiber unitaries. Each polarization measurement is considered as one compensation step.}
    \label{fig:flow_theory}
\end{figure}

In optical fibers, random birefringence caused by thermal fluctuations as well as twisting and bending the fiber change the SOP of the light passing through the fiber. Three consecutive LCVRs at $0^\circ$, $45^\circ$, and $0^\circ$ (relative to the slow axis for each LCVR) can be used to transform an arbitrary SOP to any other SOP \cite{Zhuang1999}. This is important as it allows compensating for any unitary transformation of the SOP applied by the fiber. The Mueller matrix $\mathbf{M}_{\text{LCVRs}}$ for the three LCVRs is shown in Eq.~\eqref{eq:LCVRs} in the appendix. For the polarization compensation, the algorithm sketched in Fig.~\ref{fig:flow_theory} is used. 

In the beginning, all retardances of the LCVRs are set to some arbitrary values $\delta_{i}'$. Afterwards, the SOP $\vec{S}_{\text{meas}}$ of the light is measured. For any retardances, one can calculate the SOP of the light before the LCVRs from $\vec{S}_{\text{meas}}$. The disturbed SOP $\vec{S}_{\text{dis}}$ after propagating through a single-mode fiber is given by:
\begin{equation}
\vec{S}_{\text{dis}}=\mathbf{M}_{\text{LCVRs}}^{-1}(\delta_1',\delta_2',\delta_3')\cdot\vec{S}_{\text{meas}}.
\end{equation}
Now that $\vec{S}_{\text{dis}}$ is known, one can calculate the required retardances for the LCVRs to compensate for the polarization change in the fiber. The system of equations for the compensation can be solved to get the required compensation retardances $\delta_{i}$:
\begin{equation}
\vec{S}_{\text{undis}}=\mathbf{M}_{\text{LCVRs}}(\delta_1,\delta_2,\delta_3)\cdot\vec{S}_{\text{dis}} \label{equation_sys},
\end{equation}
where $\vec{S}_{\text{undis}}$ is the desired undisturbed polarization state. The calculated compensation retardances are then set, and the polarization is measured again. The polarization at this stage is $\vec{S}_\text{comp}$. If a $97\,\%$ fidelity threshold is satisfied, we move on to the fine-tuning stage. After each fine-tuning iteration, we measure the fine-tuned state $\vec{S}_\text{fine}$. If a $99.5\,\%$ fidelity threshold is met, the algorithm stops. These thresholds can be adjusted according to the desired accuracy.

\begin{figure}[h!]
    \centering
    \includegraphics[scale=0.53]{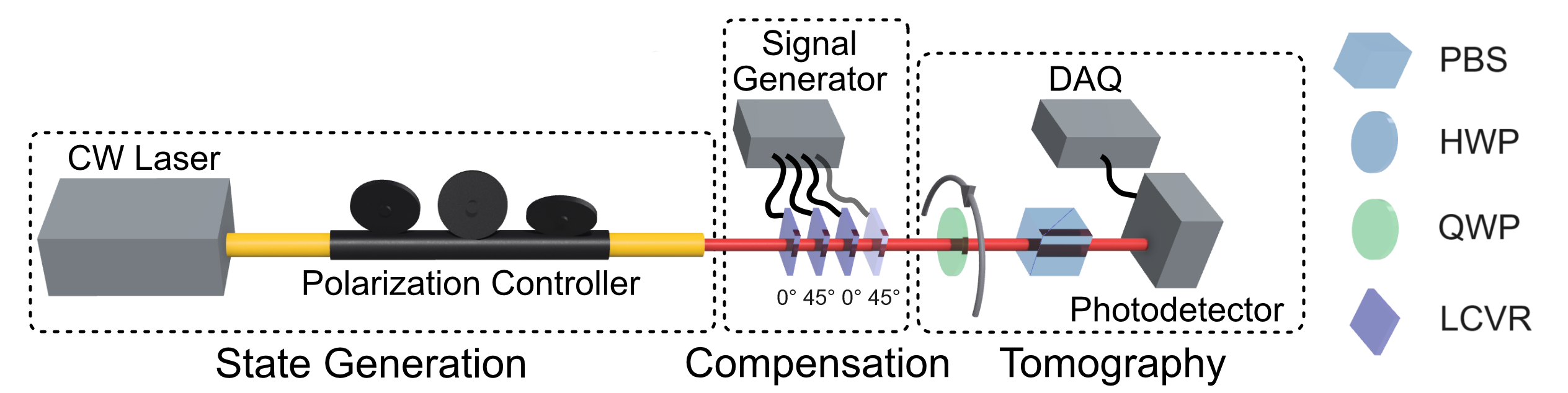}
    \caption{Setup for polarization compensation consisting of a polarization controller used to generate arbitrary disturbed input states, the LCVRs for the polarization compensation and the SOP-measurement setup; the last LCVR is optional and only used during the fine tuning.}
    \label{fig:Comp_setup}
\end{figure}

The setup in Fig.~\ref{fig:Comp_setup} was used for polarization compensation. A manual polarization controller was used to vary the disturbance of the light. The PD, the function generator, and the rotating QWP were automated and computer controlled.

Given our knowledge of $\vec{S}_\text{undis}$ and an estimate of $\vec{S}_\text{dis}$, Eq.~\ref{equation_sys} can be solved numerically using the Python \texttt{scipy.optimize.fsolve} function \cite{SciPy-NMeth} (equivalent to the analytic approach of \cite{Zhuang1999}). The values $\delta_1$, $\delta_2$, $\delta_3$ obtained from the numerical solver are not bounded and can be any real scalar. Because the LCVRs have a limited retardance range, the solution has to be shifted to be in the [$0.2\pi, 2.2\pi$) range by adding or subtracting multiples of $2\pi$. Each LCVR voltage is chosen such that its corresponding retardance is closest to the solution based on the characterization data.

Although Eq.~\eqref{equation_sys} can be solved directly (both analytically and numerically), the compensation may be less accurate if the solution is in regions where the slope of the characterization curve is steeper (Fig.~\ref{Characterization}). In that case, the LCVR cannot be controlled as precisely. A small voltage error can cause a large error in the retardance. The accuracy of the tomographic estimate $\vec{S}_\text{meas}$ also affects the compensation. These issues are addressed by returning to the step labeled ``Calculate the SOP
before the LCVRs" in Fig.~\ref{fig:flow_theory} (with different initial retardances) and repeating until the fidelity is $>97\,\%$ and a better solution is found.

Once the fidelity is $>97\,\%$, a fine tuning algorithm is used. One way to improve the result is to start with the first LCVR and make minor voltage adjustments. If fidelity improves, continue adjusting the voltage in the same direction. If not, revert to the previous voltage setting and move on to the next LCVR. Repeat this process until the $99.5\,\%$ threshold is met.

Achieving the desired fidelity requires careful calibration of the liquid crystal cells, which can be influenced by several factors. Background noise and laser power fluctuations can impact the accuracy of the calibration. Additionally, liquid crystals exhibit dependencies on temperature and wavelength, further affecting their response. These combined factors necessitate the fine-tuning of the liquid crystal cells to ensure optimal performance and achieve the target fidelity.

The system is tested by disturbing the polarization arbitrarily with a manual polarization controller in a single mode fiber (SMF) and compensating it to the arbitrarily chosen $\vec{S}_{R}$ polarization state. To prevent gimbal locking during the fine tuning \cite{shi_fibre_2021}, a fourth LCVR is added after the three LCVRs used for compensation. During the first part of the compensation, the retardance of the fourth LCVR does not change. It is only used in the fine tuning process.

\begin{table}[h!]
\centering
\begin{tabular}{ c|c|c|c } 
Test number  & Comp. steps  & Comp. steps& Comp. steps\\ 
  & until $F>97\,$\% & until $F> 99\,$\% & until $F> 99.5\,$\%\\ \hline \hline
1  & 1 & 7 & 20\\
2  & 3 & 3 & 6\\
3  & 3 & 3 & 3\\
4  & 1 & 14 & 23\\
5  & 1 & 3 & 13\\
6  & 1 & 1 & 1\\
7  & 3 & 11 & 19\\
8  & 2 & 3 & 20\\ \hline
Average  & $1.9 $ & $5.7 $ & 13.1\\
\end{tabular}
\caption{Number of compensation steps until $F>97\,$\%, $F>99\,$\% and $F>99.5\,$\%. Eight tests with different polarization controller configurations were performed.}
\label{tab: comp_result}
\end{table}

\begin{figure}[h!]
    \centering
    \includegraphics[width=0.45\textwidth]{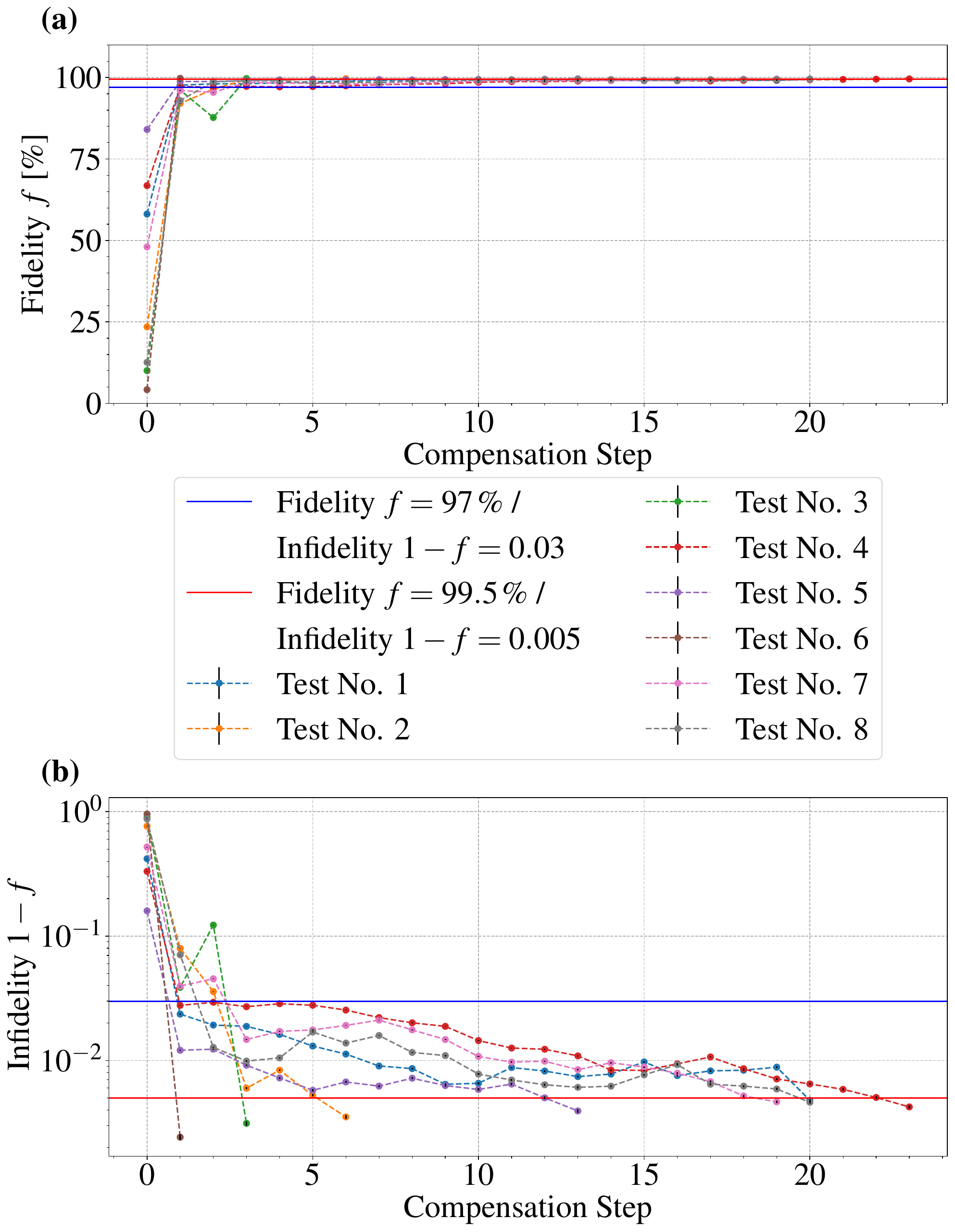}
    \caption{Polarization compensation with eight different arbitrary input states (configurations of polarization controller). (a) Fidelity plot for the full compensation process (b) Logarithmic plot of the infidelity.}
    \label{fig: comp_result_full}
\end{figure}

The compensation algorithm achieved fidelities over $97\,\%$ within an average of 1.9 compensation steps (Fig.~\ref{fig: comp_result_full} and Tab.~\ref{tab: comp_result}). Each compensation step involved measuring the SOP based on a complete QWP rotation. The subsequent fine-tuning improved accuracy until achieving fidelity above 99.5\%. Given a maximum QWP rotation speed of $25^\circ\,$s$^{-1}$, a compensation step required approximately 30 seconds (when accounting for occasional homing of the rotation mount). Therefore, six compensation steps take 180 seconds and, on average, are sufficient to achieve over 99\% fidelity. To account for variations due to different QWP rotation hardware, we chose to report our results in Fig. 8 in terms of compensation steps rather than absolute time. 

The QBER is a crucial metric in QKD that quantifies the ratio of wrong bits to the total number of bits received. According to \cite{gisin_quantum_2002}, the QBER comprises of contributions from the imperfect received state \(\text{QBER}_{\text{opt}}\), dark counts \(\text{QBER}_{\text{det}}\), and accidental coincidences \(\text{QBER}_{\text{acc}}\): 

\begin{equation}\text{QBER}=\text{QBER}_{\text{opt}} + \text{QBER}_{\text{det}} + \text{QBER}_{\text{acc}}.
\end{equation}

In polarization-based QKD, polarization compensation primarily reduces \(\text{QBER}_{\text{opt}}\) , which is the portion of the error due to unitary disturbances of the polarization state (e.g., from fiber propagation). Although Refs. \cite{gisin_quantum_2002,Anwar2021Entangled} derive this relationship in the context of entangled photon systems, we estimate the QBER using the visibility-fidelity-QBER linkage in the following way. Specifically, \(\text{QBER}_{\text{opt}}\)   can be related to the polarization visibility $V$ as:
\begin{equation}
    \text{QBER}_{\text{opt}}=\frac{1-V}{2}.
\end{equation}
and the fidelity satisfies $F\geq V$ [21]. Combining these yields the \(\text{QBER}_{\text{opt}}\) upper bound:
\begin{equation}
    \text{QBER}_{\text{opt}} = \frac{1 - f}{2}.
\end{equation}
By assuming that the fidelity of the quantum state used in QKD is on par with our measured laser-based fidelity ($F=99\%$), we estimate $\text{QBER}_{\text{opt}}\approx0.5\%$. If the combined contributions of $\text{QBER}_\text{det}$  and $\text{QBER}_\text{acc}$  are around 1\%, then the total QBER would be approximately 1.5\% after an average of 5.7 compensation iterations. This relation provides a useful approximation for estimating $\text{QBER}_\text{opt}$  based on measured polarization fidelity, offering practical insight into how improved polarization control can directly enhance QKD performance.
\section{\label{sec:V}Concluding remarks}
In conclusion, we have successfully proposed and implemented a scheme for active polarization control. Our method uses a single rotating QWP for polarization tomography, with rapid compensation achieved through LCVRs. We directly solve for the required retardances and apply further fine-tuning to reach higher fidelities. We demonstrated that our methodology can compensate for arbitrary polarization drifts within an average of fewer than six compensation iterations, achieving over $99\,\%$ fidelity.

The switching time of the LCVRs is below $150$\,ms, meaning the primary bottleneck was the rotation speed of the quarter-wave plate. Therefore, this method can achieve even quicker compensation by using a faster QWP rotation mount, such as the rotation mount used in \cite{Bobach_polarimetry_2017}. Additionally, more sophisticated fine-tuning algorithms could further increase the compensation speed.

To facilitate broader adoption, we have made the complete polarization compensation code publicly available, allowing researchers and students to implement this method with minimal effort. While our approach demonstrates robustness and efficiency, future work could focus on adapting the method to a wider range of wavelengths and urban fiber networks to improve its applicability. 

Our polarization compensation method delivers four key benefits over existing approaches. First, it provides direct, deterministic compensation via polarization tomography---bypassing time-consuming iterative or interferometric feedback and achieving $>$99\% fidelity in under six iterations, reducing dead time in applications such as QKD. Second, a built-in fine-tuning step automatically corrects for practical disturbances (e.g., LCVR temperature shifts and laser power fluctuations), improving real-world robustness. Third, we present a complete implementation from tomography through active manipulation, supplying all code and technical details to enable immediate adoption. Finally, the use of a single rotating quarter-wave plate plus LCVRs yields a minimal, integrated hardware footprint that lowers system complexity while enhancing stability against environmental perturbations.

Beyond polarization-based QKD, our technique is equally relevant to other technologies. For instance, a recent review by He \textit{et al.} \cite{He2021} highlights the growing role of polarization optics in biomedical and clinical applications, where precise polarization control is pivotal for enhancing imaging resolution and contrast. Moreover, as demonstrated by Intaravanne \textit{et al.} \cite{Intaravanne2022}, advanced manipulation of three-dimensional polarization structures can be leveraged for quantum imaging, information encoding, and integrated photonic devices, where precisely tailored input polarization is crucial.  Overall, this work provides a versatile, high-performance, and potentially scalable solution for polarization control, addressing critical challenges in modern optical systems.

\begin{acknowledgments}
We acknowledge the support of the Natural Sciences and Engineering Research Council of Canada (NSERC), Canada
Research Chairs (CRC), the Transformative Quantum Technologies Canada First Excellence Research Fund (CFREF), and the Max Planck Institute for the Science of Light in Germany.
\end{acknowledgments}

\section*{Data Availability Statement}

The data that support the results of this paper are available from the corresponding author upon reasonable request.

The code for the LCVR characterization, the polarization state tomography and the polarization compensation algorithm are available on our GitHub: \href{https://github.com/uOttawaQuantumPhotonics/polarization_compensation}{github.com/uOttawaQuantumPhotonics}.

\appendix
\section{Polarization tomography}
In this subsection of the appendix, we discuss the experimental details of the tomography intensity measurements with a PD and the angle measurement from the rotation mount. The output voltage of the PD is linearly proportional to the incoming intensity $I=k\cdot V+b$. Before starting the measurement, it is important to first measure the background noise $V_\text{back}$ without the laser turned on and subtract it from all following measurements:
\begin{equation}
    V_{n}=V^\text{raw}_{n}-V_\text{back}.
\end{equation}
Here, the index $n$ refers to the sample number in the tomography dataset, while the superscript `raw' represents the raw voltage measurement recorded by the photodetector during each polarization measurement.
Since the PD intensity is proportional to the voltage, the calculated parameters are now $A',B',C'$ and $D'$. These parameters are linearly proportional to $A_0,B_0,C_0,$ and $ D_0$, respectively, with coefficient $k$. This issue is addressed by the normalization of the Stokes parameters (see Eq.~\eqref{eq: tomography_voltage 1}-Eq.~\eqref{eq: tomography_voltage 2}). Our polarization compensation is a unitary transformation which cannot compensate for loss of degree of polarization (mixed states). This is not a significant issue, since single-mode optical fibers do not generally affect the degree of polarization, as was confirmed by our polarization tomography. Therefore, we base the compensation on the fidelity (Eq.~\eqref{eq:fidelity}), which depends only on the Stokes parameters $S_{1}$ to $S_{3}$. We can normalize the Stokes parameters by $\sqrt{S_{1}^{2}+S_{2}^{2}+S_{3}^{2}}$, where we have assumed a pure polarization state.\\
There is an offset $\alpha$ between the QWP fast axis and the $0^\circ$ on the rotation mount. The angle $\alpha$ needs to be subtracted from the measured angle $\phi$ from the motorized rotation mount as $2n\Delta\phi=\phi(2n)-\alpha$, and respectively $4n\Delta\phi=\phi(4n)-\alpha$, resulting in:
\begin{align}
\label{eq: tomography_voltage 1}
A_0&=\frac{2}{N}\sum_{n=1}^N (I^\text{raw}_n-I_\text{back}) \nonumber\\
&= k\frac{2}{N}\sum_{n=1}^N (V^\text{raw}_n-V_\text{back})=kA',\\
B_0&=\frac{4}{N}\sum_{n=1}^N (I^\text{raw}_n-I_\text{back}) \sin(\phi(2n)-\alpha ) \nonumber \\
&=k\frac{4}{N}\sum_{n=1}^N (V^\text{raw}_n-V_\text{back}) \sin(\phi(2n)-\alpha )=kB',\\
C_0&=\frac{4}{N}\sum_{n=1}^N (I^\text{raw}_n-I_\text{back})\cos(\phi(4n)-\alpha ) \nonumber\\
&=k\frac{4}{N}\sum_{n=1}^N (V^\text{raw}_n-V_\text{back})\cos(\phi(4n)-\alpha)=kC', \\
D_0&=\frac{4}{N}\sum_{n=1}^N (I^\text{raw}_n-I_\text{back})\sin(\phi(4n)-\alpha) \nonumber\\
&=k\frac{4}{N}\sum_{n=1}^N (V^\text{raw}_n-V_\text{back})\sin(\phi(4n)-\alpha)=kD', \\
\mathsf{S}_{i}' &= \frac{k\cdot S'_i}{k\cdot \sqrt{{S'_{1}}^{2}+{S'_{2}}^{2}+{S'_{3}}^{2}}} = \frac{S_i}{\sqrt{{S_{1}}^{2}+{S_{2}}^{2}+{S_{3}}^{2}}}=\mathsf{S}_{i}.
\label{eq: tomography_voltage 2}
\end{align}
Here, $S'_i$ are the Stokes parameters calculated from $A',B',C'$ and $D'$. $\mathsf{S}_{i}'$ are the normalized $S'_i$ parameters and $\mathsf{S}_i$ are the correctly normalized Stokes parameters. \\

\section{Polarization transformation}
In this subsection of the Appendix, we discuss the experimental details for the LCVR characterization and the matrix for three LCVRs. As in case of the polarization tomography, the background needs to be subtracted from the measurement of the LCVR characterization.
For each applied voltage to the LCVR, ten samples were measured by the PD. The value of one measurement $V_\text{meas}$ is the mean value of the ten samples.
The measured voltage of the PD is proportional to the input optical power: $I=k\cdot V + b $. Therefore, the retardance is given by:
\begin{align}
\delta &= \arccos\left(1 - \frac{2\cdot (V_\text{meas} - V_\text{back})}{V_\text{max} - V_\text{back}}\right),
\end{align}
which is equal to the retardance equation in Eq.~\eqref{eq: retardance} as \(k\) and \(b\) respectively cancel each other out. The uncertainty for the LCVR characterization is calculated through Gaussian error propagation and results in 
\begin{align}
\Delta \delta&=\sqrt{\left(\frac{\partial \delta}{\partial V_\text{meas}}\Delta V_\text{meas}\right)^2+\left(\frac{\partial \delta}{\partial V_\text{back}}\Delta V_\text{back}\right)^2}\\
&=\sqrt{\frac{\Delta V_\text{meas}^2+\Delta V_\text{back}^2}{(V_\text{meas}-V_\text{back})(V_{\mathrm{max}}-V_\text{meas})}}.
\end{align}\\
Here, $\Delta V_\text{meas}$ and $\Delta V_\text{back}$ represent the errors in the measured voltages and the background measurement, respectively, and are estimated using the standard error of the mean for each case. The error associated with $V_\text{max}$ is excluded from the error propagation analysis.\\

\begin{widetext}
The matrix describing the polarization transformation caused by the LCVRs is 
\begin{align}
\label{eq:LCVRs}
&\mathbf{M}_{\text{LCVRs}}(\delta_1,\delta_2,\delta_3) = 
\begin{bmatrix}
1 & 0 & 0 & 0 \\
0 & \cos\delta_2 & \sin\delta_1 \sin\delta_2 & -\cos\delta_1 \sin\delta_2\\
0 & \sin\delta_2 \sin\delta_3 & \cos\delta_1 \cos\delta_3-\cos\delta_2 \sin\delta_1 \sin\delta_3 & \cos\delta_3 \sin\delta_1+\cos\delta_1 \cos\delta_2 \sin\delta_3\\
0 & \cos\delta_3 \sin\delta_2 & -\cos\delta_2 \cos\delta_3 \sin\delta_1-\cos\delta_1 \sin\delta_3 & -\sin\delta_1 \sin\delta_3+\cos\delta_1 \cos\delta_2 \cos\delta_3 \\
\end{bmatrix},
\end{align}
\end{widetext}
\noindent where $\delta_i$ is the retardance of the $i^\text{th}$ LCVR.\\

\section{Apparatus Specification}

For the polarization tomography, we used a Thorlabs LP808-SA60 808\,nm CW laser with a $\sim1$\,nm bandwidth, a Thorlabs PDA20CS photo detector that was read out using a National Instruments 782258-01 USB-6361 DAQ, and a Thorlabs PRM1Z8 motorized rotation stage. The Thorlabs LCC1413-B compensated full-wave LCVRs were driven by JDS6600 two-channel signal generators. All components used in this work are commercially available. 

\nocite{*}
\bibliography{Ref}

\end{document}